\documentstyle[preprint,aps,epsfig]{revtex}
\def\to{\rightarrow}

\begin{document}

\preprint{
KAIST-TH-96/21}

\vspace{2.0cm}

\title{Determination of $|V_{cb}|$ from the polarization of vector meson
  in the semileptonic decay of  $B$ and  $B_c$ meson}

\vspace{2.5cm}

\author{Myoung-Taek Choi\thanks{Present address: Physics Dept., Hanyang
  University, Seoul 133-791, Korea.
  E-mail: ~mtchoi@hepth.hanyang.ac.kr}
and Jae Kwan Kim}
\vskip 1.0cm
\address{
  Department of Physics, 
  Korea Advanced Institute of Science and Technology,\\
  Taejon, 305--701, Korea}
\date{\today}
\maketitle

\begin{abstract}

Since the degree of polarization of the vector particle in semileptonic
decay strongly influences 
the decay width of the  particles,
it can be used as a measure of 
Cabibbo-Kobayashi-Maskawa quark-mixing matrix elements.
We show that $|V_{cb}|$ can be determined 
 from the measurement of
polarization of vector meson in $\bar B,\bar B_c \to Vl\bar\nu$ decay, 
where $V$ is vector meson.
\end{abstract}
\pacs{ }

\vspace{1cm}

Exclusive semileptonic decays of hadrons containing a bottom quark
provide a means to measure the Cabibbo-Kobayashi-Maskawa(CKM) matrix
elements~\cite{ckm}. 
The decay $\bar B \to D^*l\bar\nu$ is a key process for these studies.
Neubert proposed the method of model independent determination of
$|V_{cb}|$ from $\bar B \to D^*l\bar\nu$ decay
within the framework of Heavy Quark Effective Theory
(HQET)~\cite{Neubert91}. 
He observed that the zero recoil point is especially suitable for
the extraction of $|V_{cb}|$.
His method basically relies on 
the existence of one universal form factor and the fact that
the form factor is unaffected from
$1/m_Q ~(Q=b,c)$ corrections at zero recoil 
by Luke's theorem~\cite{Luke}.

The decays of the $\bar B_c$ meson are another source of
measurement of CKM element. 
The $\bar B_c \to J/\psi l \bar\nu \to (l^+l^-) l \bar\nu$ decay
is, in particular, expected to be a quite suitable mode to measure 
$|V_{cb}|$ due to  the experimentally clean signature.
Based on the fact that the matrix element of $\bar B_c \to J/\psi l
\bar\nu$ decay is expressed by only
one form factor near zero recoil point due to the spin symmetry of
heavy mesons,
Sanchis \textit{et al.}~\cite{Sanchis95}
discussed similar procedure as Neubert to extract $|V_{cb}|$.

In the above method, 
the value of $|V_{cb}|$ is extracted from the recoil
spectrum of the semileptonic decay, 
hence complete kinematic reconstruction is required.
The decay of neutral $\bar B$ meson  has been used 
because the flight direction of $\bar B^0$ is unaffected by the magnetic
field. 
In ARGUS and CLEO experiments \cite{arguspol,cleopol,cleopol2},
where  $\bar B^0$ is produced almost at rest,  
the recoil momentum of $D^{*+}$ to the $\bar B^0$ meson  
is easily calculated by reconstructing $D^0$ and $\pi^+$. 
The problem comes from the small released energy of decay products of
$D^{*+}$; since the momentum of pion of the decay is small, the
efficiency for reconstructing the charged pion is low, and the
combinatorial background for neutral pions from $D^{*0}$ decays
is significant.
In ALEPH experiment \cite{alephVcb}, the $\bar B^0$ meson is largely
boosted. The momentum  of $\bar\nu$ is calculated from the missing energy
in the hemisphere containing the $D^{*+} l$ candidate.
Although the momentum of the pion from the $D^{*+}$ decay has large
momentum,  the momentum of $\bar B^0$ meson should be reconstructed on
an event by event basis: the flight direction of the $\bar B^0$ meson is
measured by reconstructing its production and decay points.
There are uncertainties from the  missing energy
measurement and identification of particles. 

In this paper, we show that the polarization of the
vector particle can be used as a measure of the CKM element.
It is based on the fact that
the degree of polarization of the vector particle strongly influences
the momentum spectrum of the decay particles as well as the decay
rate. 
One advantage of the polarization measurement is no need of full
kinematic reconstruction: the polarization of the vector meson is
measured  only from their decay products regardless of 
the charge of the original meson($B$ or $B_c^+$).

The differential decay rate for $\bar B\to D^*l\bar\nu$ is 
\begin{eqnarray}
  \label{eq:dgam}
  \frac{1}{\sqrt{y^2-1}}\frac{d\Gamma}{dy} = \frac{G_F^2}{48\pi^3}
  m_{D^*}^3 (m_B - m_{D^*})^2 \eta_{A_1}^2 |V_{cb}|^2 \xi^2(y) \\ \nonumber
  \times 4y(y+1) \frac{1-2yr+r^2}{(1-r)^2},  ~~~~~\bar B\to D_T^*l\bar\nu, \\
  \times (y+1)^2,  ~~~~~\bar B\to D_L^*l\bar\nu.
\end{eqnarray}
Here the variable $y$ is
\begin{equation}
  \label{eq:ydef}
  y=v\cdot v' = \frac{m_B^2+m_{D^*}^2 - q^2}{2 m_B m_{D^*}}
\end{equation}
and $r=m_{D^*}/m_B$, $q^2$ the momentum transfer to the lepton pair,
$\eta_{A_1}$ the  QCD correction factor.

The form factor at non-zero recoil point is not predicted from HQET
itself, and can be parameterized by various form. Here we compare
three different functional form, linear, exponential, and the pole
ansatz~\cite{Neubert91}; 
\begin{eqnarray}
  \label{eq:xiy}
  \xi(y) &=& \xi(1)(1 - \bar\rho^2(y-1) ), \\
  \xi(y) &=& \xi(1) e^{ - \bar\rho^2(y-1)}, \\
  \xi(y) &=& \xi(1) (\frac{2}{y+1})^{2\bar\rho^2}
\end{eqnarray}
where  $\xi(1) = 1 + \delta(1/m_Q^2)$.
Then $|V_{cb}| \xi (v\cdot v')$ is fitted to recoil spectrum with
$|V_{cb}|$ and $\bar\rho$ as free parameters in the Neubert's method.

The polarization of $D^*$ is measured from the angular distribution of
their decay products in the rest frame of  $D^*$.
In $\bar B\to D^*l\bar\nu$ decay where $D^*$ decays into two
pseudo-scalar particle  $D$ and $\pi$,
the decay amplitude with the helicity $\lambda$ of the $D^*$  is
given~\cite{hagi89} 
\begin{equation}
  \label{eq:vpp}
  {\cal M} \propto \frac{G_F}{\sqrt{2}} V_{cb} \sum_{\lambda}
  L_\lambda H_\lambda Y_\lambda^1,
\end{equation}
where $\lambda=0,\pm 1$ is the helicity of the $D^*$, and hence also
that of the virtual $W^-$ in the limit of massless leptons.
$L_\lambda$ describe the $W_\lambda^{*-} \to l\bar\nu$ decay,
$H_\lambda $ are the  $\bar B \to D_\lambda^* W_\lambda^{*-}$
decay amplitudes and the $J=1$ spherical harmonics  $Y^1_\lambda$ 
describe the $D_\lambda^* \to D\pi$ decay.

The angular distribution of $\pi$ meson in rest frame of $D^*$ is
given by
\begin{equation}
  \frac{d\Gamma}{d \cos\theta_\pi} \propto 1 + \alpha \cos^2\theta_\pi, 
\end{equation}
where $\theta_\pi$ is the angle between the $\pi$ direction in the $D^*$
rest frame and the $D^*$ direction in the $B$ rest frame.
The polarization parameter $\alpha$ is related to the ratio of 
longitudinal and transverse decay width;
\begin{equation}
  \label{eq:alpha}
  \alpha = 2 ~\frac{\Gamma_L}{\Gamma_T} -1.
\end{equation}
We note that $|V_{cb}|$ and QCD
correction factor are canceled in the Eq.~(\ref{eq:alpha}).
The only unknown parameter is $\bar\rho$. 
Fitting the distribution of $D$ or $\pi$ in the rest frame of $D^*$
gives the $\alpha$, and henceforth $\bar\rho$.
Once the integrated decay rate of Eq.(\ref{eq:dgam}) is measured, 
the value of $|V_{cb}|$ is obtained.

The current measurements of polarization give 
\begin{equation}
  \label{eq:poldata}
  \alpha = \left\{
      \begin{array}{c}
        0.65\pm 0.66\pm 0.25  ~\cite{arguspol}   \\ 
        1.48\pm 0.32\pm 0.14  ~\cite{cleopol}    \\
         1.1\pm 0.4\pm  0.2   ~\cite{cleopol2}. 
      \end{array}\right.
\end{equation}
Taking the average value of these data, we obtain
\begin{equation}
  \alpha = 1.24 \pm 0.25.
\end{equation}

We show the plot of the $|V_{cb}|$ versus $\alpha$ in Figure 1, using
the linear parameterization, Eq~(\ref{eq:xiy}).
We obtain~\footnote{The error in $\bar\rho$ is
  asymmetric. This is due to the functional form of $\bar\rho$ with
  respect to the polarization in Eq.~(\ref{eq:alpha}). However, the
  $|V_{cb}|$ is  mostly symmetric as seen in Fig. 1. }
\begin{eqnarray}
|V_{cb}| &=& 0.036 \pm 0.005 \\
\bar\rho &=& 0.76^{+0.26}_{-0.53}.
\end{eqnarray}
Different ansatz produce correspondingly slight different values of
$\bar\rho$ and $|V_{cb}|$. 
The exponential ansatz leads to $|V_{cb}| = 0.036 \pm 0.006$ and
$\bar\rho=0.84^{+0.38}_{-0.60}$, while the pole gives
$|V_{cb}| = 0.037 \pm 0.007$ and $\bar\rho=0.89^{+0.40}_{-0.64}$.
Here we used~\cite{pdg96}
$\tau_B =1.56$ ps, Br$(\bar B^0 \to D^*l\bar\nu) = 4.56~\%$
to obtain $\Gamma_{\bar B^0 \to D^*l\bar\nu} = 19.2\times
10^{-15}$ GeV, and
$\eta_{A_1}\xi(1)=0.91$~\cite{Neub92}.
The uncertainties of $\tau_B$, Br($\bar B^0 \to D^*l\bar\nu$),
and $\eta_{A_1}$  which are much smaller than that of the polarization
are not taken into account in the evaluation of $|V_{cb}|$.
We note that the averaged value of $\bar\rho$ observed experimentally,
based on the fit to the recoil spectrum is  $\bar\rho = 0.93 \pm
0.06$~\cite{rho_meas}. 
The large error in our method comes from the large experimental
uncertainty in the measurement of $\alpha$. 

Now we discuss the decay of $\bar B_c$ meson. 
The $\bar B_c\to J/\psi l\bar\nu$ decay, followed by $J/\psi\to l^+l^-$,
is the most promising mode to study the semileptonic decay
of the  $\bar B_c$  meson
due to its clean  signature.
Hence the polarization of $J/\psi$ meson might be measured 
with small experimental uncertainties.

Near zero recoil point the matrix element is expressed only one form
factor~\cite{Sanchis95,Jenk93}; 
\begin{equation}
  \label{eq:meBc}
  \langle J/\psi,v' \vert A_{\mu} \vert \bar B_c,v\rangle = 
  2 \sqrt{m_{B_c}m_{\eta_c}}~ \Delta(v\cdot v') ~\epsilon_{\mu}^*.
\end{equation}
where $\epsilon_{\mu}$ is the polarization vector of the $J/\psi$.

In the limit of vanishing lepton mass, one finds;
\begin{eqnarray}
  \frac{d\Gamma}{dy} = \frac{G^2_F}{48 \pi^3}\vert V_{cb} \vert^2
  m^{2}_{B_c} m^3_{J/\psi} (y^2-1)^{1/2}(y + 1)^2 F(r,y)
\end{eqnarray}
with
\begin{eqnarray}
  F(r,y)~ =& 8(1-2y r +r^2)\left
   (\frac{\Delta}{y+1}\right)^2, 
   &\bar B_c \rightarrow J/\psi_T ~\mu~\bar\nu_{\mu}, \\
  F(r,y)~ =& 4(y-r)^2 \left (\frac{\Delta}{y+1}\right)^2, 
 &\bar B_c \rightarrow J/\psi_L ~\mu~\bar\nu_{\mu},
\end{eqnarray}
where  the corresponding variable $y$ should be used instead
of Eq.~(\ref{eq:ydef}).

Using operator product expansion, Jenkins \textit{et
  al.}~\cite{Jenk93}  showed that  
$\Delta$ can be expressed by wave function overlaps
of initial and final meson in the zero recoil limit:
\begin{equation}
  \Delta(1) = \int d^3\vec{x} ~\Psi^*_{J/\psi}(\vec{x})~\Psi_{\bar
  B_c}(\vec{x}), 
\end{equation}
which is the wave function overlap of initial and final hadron.
Because the flavour symmetry is broken, $\Delta(v\cdot v')$ is not
normalized to 1 at zero recoil point.
The wave function overlap might be calculated from the 
lattice QCD calculation or quark potential models in a reliable way
because of non-relativistic nature of the $\bar B_c$ meson.

Deviations from the spin symmetry imply the appearance
of new form factors in the hadronic matrix element which
do not contribute at zero recoil. 
Spin symmetry breaking effects can occur when the $c$-quark's
recoil momentum is larger than $m_c$. 
However we expect  that
the Eq.~(\ref{eq:meBc}) is applicable to other kinematic point
since the recoil momentum of $J/\psi$ is small ($y_{\mbox{max}}-1
\simeq 0.26)$ due to its heavy mass.

In $\bar B_c \to J/\psi l \bar\nu$ decay
where  the $J/\psi$ meson decays into two
lepton pair,
the decay amplitude with the helicity $\lambda$ of the $J/\psi$ meson is
obtained in a similar way to Eq.~(\ref{eq:vpp}),
corresponding $Y_\lambda^1$ be replaced by Wigner $d$-function
\begin{equation}
  \label{eq:vll}
  {\cal M} \propto \frac{G_F}{\sqrt{2}} V_{cb} \sum_{\lambda}
  L_\lambda H_\lambda d^1_{\lambda_{J/\psi},\lambda_{l^-}-\lambda_{l^+}}, 
\end{equation}
where $\lambda_{l^-}-\lambda_{l^+} = \pm 1$.

The angle distribution of $l^-$ in rest frame of $J/\psi$ is given
\begin{equation}
  \frac{d\Gamma}{d \cos\theta_{l^-}} \propto 1 + \alpha'
  \cos^2\theta_{l^-},  
\end{equation}
where we find
\begin{equation}
  \alpha' = \frac{\Gamma_T - 2\Gamma_L}{\Gamma_T + 2\Gamma_L}.
\end{equation}

The  parameter $\alpha'$ is again related to the ratio of 
longitudinal and transverse decay widths.
With the observation that 
the  $\Delta(1)$ is canceled from the expression, 
one obtains $\bar\rho$ from the  polarization of the $J/\psi$ meson.
Since unknown two parameters are obtained~\footnote{We note that
one should be careful to 
parameterize the form factor $\Delta(y)$
in the linear form like Eq.(\ref{eq:xiy}) because $\bar\rho^2$
is predicted to be large, even $2 \sim 4$ in ISGW model~\cite{isgw}},
one can calculate $|V_{cb}|$
once the integrated decay width is measured, 
as in $B \to D^* l \bar\nu$ case.

In summary,  we discussed an alternate way to extract $|V_{cb}|$
from the measurement of polarization of vector meson in the
exclusive semileptonic decays. 
The  advantage of the method is that the full kinematic
reconstruction is not needed in this case.

Full use of kinematic information 
such as angular correlations of the decay products
is expected to allow a determination of the invariant
hadronic form factors and the $|V_{cb}|$ quark mixing matrix element
almost model-independently~\cite{hagi89}.
The semileptonic decays of $B_c$ meson is quite suitable for those study
due to the clean signature.

\begin{center}
  \bf Acknowledgements
\end{center}

This work was supported 
in part by the Korea Science and Engineering Foundation.


\newpage
\begin{figure}
  \begin{center}
    \caption{$|V_{cb}|$ versus $\alpha$ in $\bar B\to
     D^*l\bar\nu$ (using the linear parametrization of the form factor). 
     $\tau_B = 1.56 \pm 0.06$ ps, 
     Br($\bar B\to D^*l\bar\nu) = 4.56 \pm 0.27~\% $ is used}
  \end{center}
\end{figure}

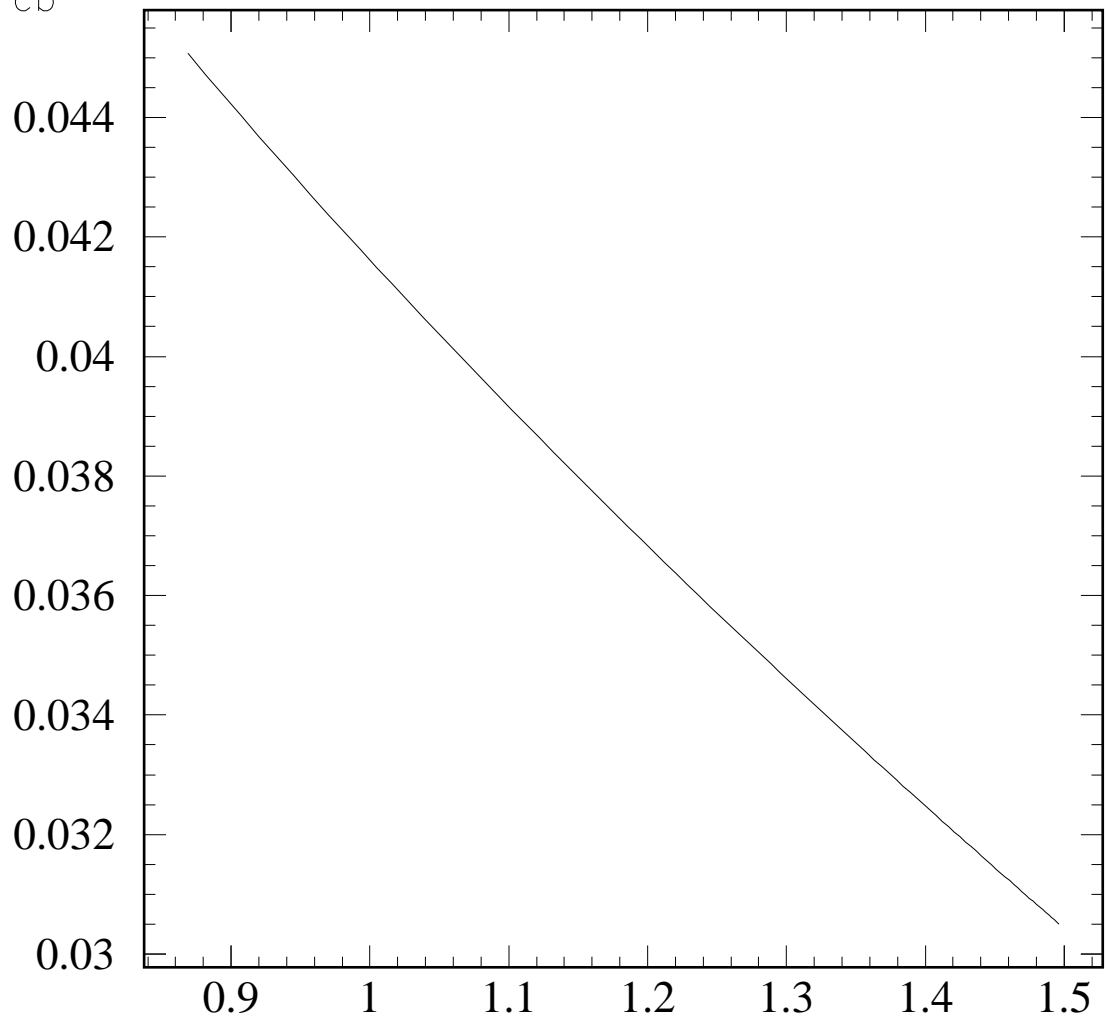
\end{document}